\documentclass[a4paper,twoside]{article}

\usepackage[authoryear,square]{natbib}
\bibpunct{(}{)}{;}{a}{}{,}
\usepackage{psfig}
\usepackage{times}
\date{} %Please leave the date blank
\newcommand{\HI}{H\,{\sc i}}
\newcommand{\HII}{H\,{\sc ii}}

\newcommand{\etal}{et al.}
\def\ga{\ifmmode\stackrel{>}{_{\sim}}\else$\stackrel{>}{_{\sim}}$\fi} 
\def\la{\ifmmode\stackrel{<}{_{\sim}}\else$\stackrel{<}{_{\sim}}$\fi} 
\title{\large\bf\flushleft The Vertical Structure of Warm
Ionised Gas in the Milky Way}
\author{\parbox{\textwidth}{\flushleft
\vspace{-0.5cm}
{B. M. Gaensler,$^{A,D}$
G. J. Madsen,$^{A,B}$, S. Chatterjee$^A$ and S. A. Mao$^C$}\\
\vspace{0.4cm}
{\small $^A$Institute of Astronomy, School of Physics, The University of Sydney, NSW 2006,
Australia}\\
{\small $^B$Department of Astronomy, University of Wisconsin,
Madison WI 53706, USA}\\
{\small $^C$Harvard-Smithsonian Center for Astrophysics, Cambridge MA 02138,
USA}\\
{\small $^D$ARC Federation Fellow; email: bgaensler@usyd.edu.au}}}
\begin{document}

\twocolumn[
\begin{changemargin}{.8cm}{.5cm}
\begin{minipage}{.9\textwidth}
\vspace{-1cm}
\maketitle

%%%%%%%%%%%%%     ABSTRACT    %%%%%%%%%%%%%
%sof no more than 200 words here.
\small{\bf Abstract:} 
We present a new joint analysis of pulsar dispersion measures and
diffuse H$\alpha$ emission in the Milky Way, which we use to derive
the density, pressure and filling factor of the thick disk component
of the warm ionised medium (WIM) as a function of height above the
Galactic disk. By excluding sightlines at low Galactic latitude
that are contaminated by \HII\ regions and spiral arms, we find
that the exponential scale-height of free electrons in the diffuse
WIM is $1830^{+120}_{-250}$~pc, a factor of two larger than has
been derived in previous studies. The corresponding inconsistent
scale heights for dispersion measure and emission measure imply
that the vertical profiles of mass and pressure in the WIM are
decoupled, and that the filling factor of WIM clouds is a geometric
response to the competing environmental influences of thermal and
non-thermal processes. Extrapolating the properties of the thick-disk
WIM to mid-plane, we infer a volume-averaged electron density
$0.014\pm0.001$~cm$^{-3}$, produced by clouds of typical electron
density $0.34\pm0.06$~cm$^{-3}$ with a volume filling factor
$0.04\pm0.01$. As one moves off the plane, the filling factor
increases to a maximum of $\sim30\%$ at a height of $\approx$1--1.5~kpc,
before then declining to accommodate the increasing presence of
hot, coronal gas.  Since models for the WIM with a $\approx1$~kpc
scale-height have been widely used to estimate distances to radio
pulsars, our revised parameters suggest that the distances to many
high-latitude pulsars have been substantially underestimated.

%%%%%%%%%%%%%     KEYWORDS    %%%%%%%%%%%%%
\medskip{\bf Keywords:} 
galaxies: ISM ---
Galaxy: halo, structure ---
globular clusters: general ---
ISM: structure ---
pulsars: general
% Please write all keywords in lower case. PASA uses the
% standard list of subject headings adopted by The Astrophysical Journal
% and available from http://www.journals.uchicago.edu/ApJ/keywords_text.html.
% Keywords are separated by em-dashes, i.e. ---

%%%%%%%%DO NOT EDIT%%%%%%%%%%%%
\medskip
\medskip
\end{minipage}
\end{changemargin}
]
\small
%%%%%%%%EDIT FROM HERE%%%%%%%%%%%%

\section{Introduction}
\label{sec_intro}

The interstellar medium (ISM) of the Milky Way is a complex,
multi-phase environment. Much of the ISM is an ionised gas at
temperatures of $10^4-10^6$~K, requiring one or more vast sources
of ongoing energy injection. In the modern picture of the ISM
\citep[see][for a review]{fer01}, ionised gas consists of two main
phases: a hot ``coronal'' component at temperatures
of $\sim10^5-10^6$~K, and a warm ionised medium (WIM) of temperature
$\sim10^4$~K.

The WIM is most easily identified in H$\alpha$ and other optical
recombination lines, which show that much of this gas corresponds 
to \HII\ regions
around massive stars. However, most \HII\
regions are seen only at low Galactic latitudes. Further from the plane,
pulsar dispersion measures (DMs), free-free absorption of low-frequency
Galactic synchrotron emission, interstellar scattering of compact radio
sources, and faint H$\alpha$ and H$\beta$ emission all demonstrate the
existence of a widespread, diffuse WIM, not associated with individual
stars \citep[see overviews by][]{kh87,rey90c}.  In the rest of this
paper, we use WIM to refer only to this diffuse component, disregarding
individual \HII\ regions.

The diffuse WIM has a typical volume-averaged free electron density
$n \approx 0.01-0.1$ cm$^{-3}$ and an electron temperature $T_e
\approx 8000$~K \citep[e.g.,][]{rey90c,wsx+08}.  Scattering and
dispersion of high-latitude pulsars demonstrate that the diffuse
component of the WIM is distributed in a thick disk, extending more
than a kpc above and below the Galactic plane with a roughly
exponential fall-off in density \citep{rd75,rey89,nct92b}. The power
required to maintain the ionisation state of the thick-disk WIM is
$\sim5\times10^{41}$~ergs~s$^{-1}$, a vast rate of energy injection
that can seemingly only be supplied by the photo-ionising flux of
all the hot stars in the Galaxy (Reynolds \etal\ 1984,
1990b\nocite{rey84,rey90b}). Questions remain, however, over how
these ultraviolet photons propagate to large distances above the
Galactic plane \citep{rey90,mc93}, and whether the spectrum of the
photon field produced by massive stars is consistent with the
observed line ratios in the WIM \citep{rt95b,hklr96,mrh06}.  Many
other edge-on spiral galaxies also have a faint, thick WIM disk,
seen in deep H$\alpha$ observations \citep{ran98,det04}.  All these
results make clear that the WIM is a key part of the feedback process
between the stellar and gaseous components of a galaxy, and that
it connects the energetics and dynamics of the disk and the halo.
There is thus considerable motivation to understand the density,
spatial distribution and extent of the WIM in the Milky Way and in
other galaxies.

Our own Galaxy is unique in that
the DMs of radio pulsars can provide the free-electron column of the WIM for thousands
of lines of sight, allowing one to develop a smooth underlying model
for the three-dimensional Galactic distribution of $n$
\citep[e.g.,][]{tc93,gbc01}.  Integrating such a distribution allows
one to also predict the emission measure (EM) along any sightline.
However, even when interstellar extinction is taken into account,
the predicted EMs are vastly smaller than those inferred from diffuse
H$\alpha$ emission, from low-frequency absorption in the spectra
of non-thermal radio sources, or from free-free radio emission
\citep[e.g.,][]{pw02,srwe08}.  
Because DMs have a linear dependence on the free electron
density while EMs depend on the square of the electron density,
the inference is that even far from
the plane, the WIM is not a smooth, homogeneous medium, but is
clumped into discrete clouds or layers \citep{rey77,rey91,pyn93}.
In the following sections, we review some recent attempts
to model this structure.

\subsection{Modeling The WIM}

\subsubsection{The NE2001 Model}
\label{sec_ne2001}

Recognising the limitations of a smooth, homogeneous WIM, Cordes \&
Lazio (2002, 2003\nocite{cl02,cl03}) developed a detailed model, termed ``NE2001'',
for the structure of ionised gas in the Galaxy.
The NE2001 model consists of a thin disk of scale height 140~pc
associated with low-latitude \HII\ regions; a thick-disk WIM layer as
discussed above, with a scale height of 950~pc; large-scale structure
associated with spiral arms; cavities and density enhancements
corresponding to known features in the local ISM; and individual
clumps and voids needed to model specific regions of enhanced/reduced
scattering or dispersion.  The NE2001 model can predict the DM, EM,
angular broadening, scintillation bandwidth and other parameters
for any sightline integrated to any distance. It is widely used to
estimate distances to radio pulsars for which there is no other distance
indicator other than their DMs.

The NE2001 model attempts to provide a comprehensive description
of the WIM, and can be continually improved as more DMs and other
data are obtained. However, as \cite{cl02} acknowledge, the NE2001
model fails to correctly predict the DMs of some pulsars in
high-latitude globular clusters.  For example, the 22 pulsars in
the globular cluster 47~Tuc have an average DM of 24.4~pc~cm$^{-3}$
and are at a distance of $\approx5$~kpc \citep{clf+00,gbc+03}.
Pulsars in this cluster are included as an input to the NE2001
model, but the prediction of NE2001 is DM~$=42.1$~pc~cm$^{-3}$, in
significant disagreement with the observations.  More broadly,
\cite{lfl+06} have shown that if NE2001 is applied to the new pulsars
subsequently discovered in the various surveys that used the Parkes
multibeam receiver, the inferred vertical distribution of these
sources in the Galaxy has a scale height about a factor of two lower
than expected.

The NE2001 model also has other difficulties at high latitudes.  At
the north Galactic pole, NE2001 predicts\footnote{We have used the software
provided at {\tt http://www.astro.cornell.edu/$\sim$cordes/NE2001/},
v1.0.} EM~$\approx 0.05$~pc~cm$^{-6}$. This is more than order of
magnitude smaller than that observed \citep{rey91,pw02,hrt+03},
suggesting that there is substantial clumping of the WIM at large
distances from the Galactic plane, not included in NE2001 \citep[see
also discussion by][]{rey91,rey97}.

Meanwhile, \cite{srwe08} have attempted to simultaneously reconcile
extragalactic Faraday rotation measurements and all-sky maps of
Galactic radio emission with the distribution of $n$ predicted by
NE2001. They show that a joint fit to all these data requires an
anomalously strong halo magnetic field, and an unrealistically low
scale-height for Galactic cosmic rays. \cite{srwe08} argue that if
the scale-height of the thick-disk WIM in the NE2001 model were increased
from 950~pc to $\sim2$~kpc, a self-consistent model for the Galactic
distribution of thermal gas, magnetic fields and cosmic rays could
then be derived.

\subsubsection{The BMM06 Model}
\label{sec_bmm06}

The arguments in \S\ref{sec_ne2001} suggest that the scale height
and volume filling factor of the thick-disk WIM need to be re-evaluated. A
detailed study of this issue has been carried by \cite{bmm06},
hereafter BMM06, who used measurements of EM and DM for 157 pulsars
to derive statistics on the density, scale height and filling factor
of the WIM. BMM06 argued that both the internal cloud electron
density, $N$ and volume filling factor, $f$, of the clumpy WIM
evolve with a source's vertical distance, $z$, above or below the
Galactic plane. They fit the data with a model in which $N$ decays
exponentially, while $f$ grows exponentially, with $z$. The expressions
for their best fits are:
\begin{equation}
N(z) = N_0~e^{-\frac{z}{H_N}} ,~
f(z) = f_0~e^{+\frac{z}{H_f}} ,
\label{eqn_bmm1}
\end{equation}
with scale heights
\begin{equation}
H_N = 710_{-120}^{+180}~{\rm pc},~
H_f = 670_{-130}^{+200}~{\rm pc},
\label{eqn_bmm2}
\end{equation}
and extrapolated mid-plane values
\begin{equation}
N_0 = 0.41\pm0.06~{\rm cm}^{-3}, 
f_0 = 0.05\pm0.01 .
\label{eqn_bmm3}
\end{equation}

The resulting volume-averaged density of the thick-disk WIM, $n \equiv fN$, remains
roughly constant at $n \approx 0.02$~cm$^{-3}$ up to $z \sim 1$~kpc,
implying a much larger effective scale height for the WIM layer
than had been assumed by all previous authors.
At higher $z$, BMM06 argue that the filling factor takes on a
constant value $f \ga 0.3$, while $N$ continues to decay.  As
discussed by \cite{srwe08}, the NE2001 model cannot successfully
predict the free-free emission seen at 23~GHz by the {\em WMAP}\
experiment, but incorporating the
function $f(z)$ derived by BMM06 (see Eqn.~[\ref{eqn_bmm1}] above)
produces a good match to the data.

The BMM06 model clearly offers a number of improvements over previous
work in its description of the WIM at high $z$. However, there are
several important caveats and limitations associated with that
study. First, for 95\% of the pulsars used by BMM06, no independent
estimates of distance (or hence of $z$) were available, so BMM06
used the distances predicted by the NE2001 model, and assumed errors
of 20\% in these distances.  However, as we noted in \S\ref{sec_ne2001},
the NE2001 model can be in error by much more than this at high
Galactic latitudes.  
Conversely, BMM06 explicitly excluded from their sample all pulsars
in globular clusters or in the Magellanic Clouds, eliminating a
substantial number of sources with reliable, independent
distances.\footnote{A subsequent study by \cite{bm08} has repeated
the analysis of BMM06 but only using pulsars with independent
distance estimates.  However, for reasons that they do not explain, \cite{bm08}
only consider 13 of the 25 globular clusters known to contain radio
pulsars, and do not use information on pulsars in the Magellanic Clouds.}

A second limitation of BMM06 \citep[and also of the subsequent study
by][]{bm08} is that their data sets
consist of paired estimates of DM and EM for each pulsar. While the
DM is a direct integral of $n$ from the observer to the source, the
EMs are derived from H$\alpha$ data, which correspond to integrals
to infinity, attenuated by dust extinction. BMM06 scaled down the
EM values to account for the finite distance of each pulsar, but
this calculation propagates the significant distance uncertainties
in NE2001 into the EM values. Furthermore, scaling the EMs requires
one to assume a WIM scale height, introducing a degeneracy into
their subsequent fitting of the EMs to derive this same scale height.
BMM06 corrected the EMs for extinction using
standard reddening laws, but such corrections assume that dust and
gas are well mixed along the line of sight. A inhomogeneous dust
distribution, combined with significant distance uncertainties for
most pulsars, can result in an erroneous extinction correction.

Finally, BMM06 considered pulsars with $60^\circ < \ell < 360^\circ$
and $|b| > 5^\circ$ (where $\ell$ and $b$ are Galactic longitude
and latitude, respectively). As we will show in \S\ref{sec_distrib},
the DMs of many pulsars at $|b| < 40^\circ$ are contaminated
by \HII\ regions and by other low-latitude structure.  When trying to
extract the properties of the WIM in the thick disk, the DMs and
EMs of such pulsars will increase the scatter in the data and will
reduce the signal-to-noise ratio of the underlying signal.

Overall, we conclude that while the approach of BMM06 provides
several new insights into the behaviour of the WIM
as a function of $z$, there are large systematic uncertainties that
need to be incorporated into the parameters they extract from the
data.

\subsubsection{Scope Of This Paper}

The discussion in \S\ref{sec_ne2001} and \S\ref{sec_bmm06} illustrates
the strengths and weaknesses of the NE2001 and BMM06 models.
Specifically, NE2001 is a detailed description of ionised gas along
existing sightlines, but 
has limited predictive power at high
$z$ where the number of existing sightlines is sparse. BMM06 provide
a comprehensive joint analysis of EMs and DMs, and lay out a
functional form for the $z$-dependence of the thick-disk WIM that is a
significant improvement over that of NE2001. However, their analysis
suffers from the significant uncertainties introduced by adopting
distances from NE2001, and by trying to correct EMs for extinction
and for the finite path length to the pulsars in their sample.

In this paper, we complement these previous studies,
with a simple attempt to model the overall structure of the
diffuse WIM. Our main thrust is to re-apply the analysis of \cite{rey89},
who plotted DM$_\perp\equiv$ DM$.\sin|b|$ vs. $z$ for $\sim35$
pulsars with known distances, and compared this to the predictions
of simple geometric models. While this approach has been repeated
in various subsequent papers \citep[e.g.,][]{nct92b,gbc01,bm08}, the
much larger sample of pulsars now available affords us the luxury
of being able to filter the data to avoid contamination by the
clumps, spiral arms and \HII\ regions incorporated in NE2001, and
to thus boost any signal produced by the truly diffuse WIM.  In
\S\ref{sec_dist}, we review the sample of pulsars with known,
reliable, distances and define our sample of 53 sightlines.  In
\S\ref{sec_analysis} we use a cut on these DMs in Galactic latitude 
to derive a substantially revised scale-height of the thick-disk WIM,
and then
combine these DMs with separate data on
EMs (i.e., not matched to pulsar sightlines, as carried out by BMM06
and Berkhuijsen \& M\"uller 2008\nocite{bm08})
to confirm the contention of BMM06 that the data are
most simply explained if the volume filling factor
of this layer evolves with~$z$.  In \S\ref{sec_disc} we discuss
the implications of our calculations for the distances
to pulsars, for the structure of the WIM,
and for the transition from warm to million-degree gas in the
Galactic halo.

\section{Pulsars with Reliable Distance Estimates}
\label{sec_dist}

Independent distances for pulsars can be determined through a variety
of techniques.  Most commonly, pulsar distances are derived
kinematically, in which measurements are made of the systemic
velocities of foreground clouds seen in \HI\ absorption (either
against emission from the pulsar itself or against an associated
object such as a supernova remnant).  A model for Galactic rotation
can then be used to convert these velocities into upper and/or lower
limits on the pulsar's distance  (see \citealt{fw90} for a review).

While the kinematic approach produces reliable distances for many
pulsars, there is a key difficulty in using these data to study the
properties of the thick-disk component of the WIM. Since our explicit
aim is to study the diffuse component of the WIM, we wish to avoid
lines of sight that pass through more complicated regions of the
ISM. Since kinematic distances rely on the clumpy and complicated
nature of interstellar gas to produce discrete absorption features,
sightlines with good \HI\ absorption measurements are precisely
those we wish to exclude if we want to study a diffuse component
of the ISM.  We thus exclude kinematic distances from further
consideration, and below focus on three remaining samples of pulsars
with distances determined independently of the foreground ISM
properties.  In any case, even when using only these other distance
indicators, we will argue in \S\ref{sec_high} that one must exclude
lines of sight for which $|b| < 40^\circ$ to avoid contamination
by \HII\ regions and spiral structure. Since \HI\ absorption distances
are not feasible for pulsars with $|b| \ga 10^\circ$ \citep[see,
e.g.,][]{wsx+08}, this makes moot any concerns as whether \HI\
absorption distances should be included in our sample.

\subsection{Pulsars with Parallaxes}
\label{sec_px}

The most reliable pulsar distances come from parallaxes,
derived either through interferometric measurements or through 
pulsar timing. The
pulsar catalogue\footnote{On-line catalogue at \\ {\tt
http://www.atnf.csiro.au/research/pulsar/psrcat}, version 1.33, dated 2008~Jul~19.}
of \cite{mhth05} lists 34 such pulsars which are also radio emitters
(and hence have DM measurements). 
From this sample,
we then eliminate seven objects for which the uncertainty in the
parallax is greater than 33\%, and also discard the Vela pulsar
(PSR~B0833--45), because the DM of this nearby pulsar is dominated
by contributions from its associated young supernova remnant
\citep{bac74,dgr+92}.

\subsection{Pulsars in Globular Clusters}
\label{sec_glob}

Our second reliable sample are radio pulsars in globular clusters.  The
on-line compilation\footnote{{\tt
http://www.naic.edu/$\sim$pfreire/GCpsr.html}~, version dated 2008
Mar 13} of such sources maintained by Paulo Freire lists 137 pulsars
in 25 globular clusters.
The DMs of pulsars within each
cluster show a small level of variation, resulting both from ionised
gas internal to the cluster \citep{fkl+01} and from small-scale
angular fluctuations in the Galactic DM between adjacent sight lines
\citep{ss02,ran07}.  However, the fractional variation in DM for
each cluster is very small: the largest scatter in DM is for NGC~6760,
for which the fractional variation between the two detected pulsars
is $\approx2\%$. We correspondingly consider each globular cluster
as a single datum (regardless of how many pulsars the cluster
contains), and adopt the mean and standard deviation of the pulsar
DMs in that cluster as the representative value and uncertainty of
the DM, respectively, for that cluster. Distances to each cluster
have been taken from the on-line version\footnote{{\tt
http://physwww.mcmaster.ca/$\sim$harris/mwgc.dat}~, version dated
2003 Feb} of the compilation of \cite{har96}. The distances were
derived from the colour-magnitude diagram of each cluster by
identifying the mean magnitude of the horizontal branch population,
and as an ensemble should be reliable to $\approx15\%$ (C.\ Heinke,
private communication); we consequently adopt this as the fractional
distance uncertainty for each cluster.  

\subsection{Pulsars in the Magellanic Clouds}
\label{sec_mag}

Our final sample are pulsars in the Magellanic Clouds. The pulsar
catalogue of \cite{mhth05} lists 14 radio pulsars in the Large
Magellanic Cloud (LMC) and five radio pulsars in the Small Magellanic
Cloud (SMC) \citep[see also][]{ckm+01,mfl+06}. Since the SMC and LMC
themselves have significant DM contributions \citep{ghs+05,mfl+06,mgs+08},
these systems can only provide an upper limit on the Galactic DM
contribution along these sight lines. For the LMC, the lowest DM
is that of PSR~J0451--67, but there is uncertainty as to whether
this pulsar is actually in the LMC, or is a Galactic foreground
source \citep{mfl+06}. We correspondingly adopt the next highest
DM, 65.8~pc~cm$^{-3}$ for PSR~J0449--7031, as our upper limit in the direction
of the LMC.  For the SMC, the lowest DM is 70~pc~cm$^{-3}$ for PSR J0045--7042.
We assume distances to the LMC and SMC of 50 and 61~kpc, respectively
\citep[][and references therein]{wal03,alv04}.

\section{Analysis}
\label{sec_analysis}

\subsection{The Distribution of Dispersion Measures}
\label{sec_distrib}

The DM of a pulsar is defined as
\begin{equation}
{\rm DM} = \int_0^{d} n(l)~dl ,
\label{eqn_dm}
\end{equation}
where $d$ is the distance to the pulsar and $n(l)$ is the free electron 
density along a line element $dl$.
The distributions of DM$_\perp\equiv$~DM$.\sin|b|$ as a function of  $z$ for
the 51 measurements (\S\ref{sec_px} \& \S\ref{sec_glob}) and two
upper limits (\S\ref{sec_mag}) are shown in Figure~\ref{fig_dmz}.
It is immediately clear that for pulsars with vertical heights in
the range $10 \la z \la 1000$~pc, the average value of  DM$_\perp$
steadily increases with $z$, indicating that such sources sit within
the Galactic electron layer.  For heights $z \ga 1000$~pc, values
of DM$_\perp$ saturate at $\approx25-30$~pc~cm$^{-3}$, correspondingly
demonstrating that the electron layer falls off in density at these
values of $z$. This saturation value is consistent with the upper
limits for the LMC and SMC. In this sense, the data clearly reproduce
the findings of many previous studies
\citep[][BMM06]{rey89,rey91b,sed90,bv91b,nct92b,mc93,gbc01,cl03,hbrh05,bm08}.

The overall distribution of ionised gas in the thick disk is reasonably described
by a single plane-parallel layer \citep[e.g.,][]{hrt+03}.  To confirm
that our data are consistent with such a geometry, and to estimate
the scale height of this structure, we compare our measurements to
a planar distribution of free electron density, $n(z)$, which
falls off exponentially with increasing $z$:
\begin{equation}
n(z) = n_0~e^{-\frac{z}{H_n}} ,
\label{eqn_nz}
\end{equation}
where $n_0$ is the extrapolated density of ionised gas at mid-plane and $H_n$
is the corresponding scale height. Integrating
Equation~(\ref{eqn_nz}) out to a given distance then gives the
predicted distribution of DM$_\perp(z)$:
\begin{equation}
{\rm DM}_\perp(z) = n_0~H_n~\left(1 - e^{-\frac{z}{H_n}}\right) .
\label{eqn_dmz}
\end{equation}

\subsubsection{Exponential Fit To All Data}
\label{sec_all}

We have fit the 51 measurements in Figure~\ref{fig_dmz} to
Equation~(\ref{eqn_dmz}) using a Levenberg-Marquardt algorithm
\citep{ptvf92} (the two upper limits from the Magellanic Clouds are
excluded from the fit, but are used as independent tests on the
validity of any fitted model). Note that since the errors in
DM$_\perp$ are negligible compared to the errors in $z$, we neglect
the former and adopt DM$_\perp$ as our independent variable, with
each datum weighted by the inverse square of the error in $z$.

A fit of all the data to Equation~(\ref{eqn_dmz}) does not converge.
This is due to the very small errors in $z$ for pulsars with distances
from parallaxes: since there is no single curve that passes through
all these points, the resulting $\chi^2$ is always high, with no
set of fit parameters that correspond to a significant minimum.
This problem can be mitigated by recognising that although the
errors in distances to individual pulsars are indeed small, there
is an additional systematic effect, in that random clumps of electrons
along individual sightlines cause each measurement to deviate from
the smooth underlying model \citep{gbc01,cl02}.

Rather than explicitly incorporate individual clumps toward each pulsar
as for the NE2001 model,  we
represent the overall systematic limitations of the data by including
in the fit an additional fractional uncertainty (added in quadrature)
to the value of $z$ for each pulsar \citep[see, e.g.,][]{sed90}.
Adding successively larger fractional errors establishes
that an additional 10\% systematic uncertainty allows the fit to
Equation~(\ref{eqn_dmz}) to converge, with a reduced chi-squared, $\chi^2_r$,
of 12.3 for 49 degrees of freedom. 
The best-fit parameters are $H_n = 1010^{+40}_{-170}$~pc
and $n_0 = 0.031^{+0.004}_{-0.002}$~cm$^{-3}$, where the uncertainties
have been determined via a bootstrap algorithm
\citep{et93}, and correspond to the most compact 68\% interval of
bootstrap realisations around the best-fit values.  The corresponding
curve is plotted as a dashed line in Figure~\ref{fig_dmz}: this fit is
consistent with the upper limits on DM$_\perp$ for the Magellanic Clouds,
and is quite similar to the thick-disk component of NE2001, for
which $H_n = 950$~pc and $n_0 = 0.035$~cm$^{-3}$ (plotted as a
dot-dashed line in Fig.~\ref{fig_dmz}).\footnote{\protect\cite{cl02}
adopt a distribution $n(z) \propto {\rm sech}^2 (z/H_n)$ rather
than the exponential form of Eqn.~(\ref{eqn_nz}), but the two functions
are virtually indistinguishable, especially at high~$z$.} 

\subsubsection{Exponential Fit To High-Latitude Data Only}
\label{sec_high}

While the results in \S\ref{sec_all} demonstrate that we can reproduce the
findings of previous studies, the quality of the overall fit is reasonably
poor. In particular, the best-fit curve substantially over-estimates
DM$_\perp$ for most pulsars at $z > 1000$~pc. A possible cause of this
discrepancy is indicated in Figure~\ref{fig_aitoff}, where we plot
the distribution of our sample on the sky in Galactic coordinates.
This demonstrates that there is a significant concentration of sources
(mainly pulsars in globular clusters) in the Galactic plane and toward
the Galactic Centre, for which the DMs are relatively high. This is
simply understood as resulting from these pulsars being viewed not
just through the thick-disk component of the WIM, but also through
an additional thin-disk layer, corresponding to the dense, turbulent,
ionised gas in individual \HII\ regions and supershells. The greyscale
in Figure~\ref{fig_aitoff} clearly demonstrates this effect (see also
Fig.~6 of BMM06), showing that low-latitude pulsars usually lie behind
complicated regions of enhanced H$\alpha$ and EM.

We have correspondingly colour-coded the data in Figure~\ref{fig_dmz}
by Galactic latitude. It can indeed be seen that 
the red and orange points tend to lie
above the blue and green ones, demonstrating that for a given
$z$, sources at low $|b|$ tend to have higher values of DM$_\perp$
than at high $|b|$. To isolate the thick disk component, we thus
fit to successive subsets of the data, selected by adopting a minimum
threshold for $|b|$, $|b|_{\rm min}$, and thereby excluding sources
contaminated by the thin-disk component of the WIM.

The results of these fits for $0^\circ \le |b|_{\rm min} \le 75^\circ$
are shown in Figure~\ref{fig_data}.  This plot demonstrates that
the quality of the fit is relatively poor, $\chi^2_r \approx 12-13$,
for $0^\circ \le |b|_{\rm min} \le 30^\circ$, but is significantly
improved, $\chi^2_r \approx 4-5$, for fits with $40^\circ \le
|b|_{\rm min} \le 65^\circ$ (above this value, the sample size
becomes too small to be meaningfully analysed). We adopt as our
best description of ionised gas in the thick disk the lower latitude
threshold that maximises the size of the sample while maintaining
a good fit. This value is $|b|_{\rm min} = 40^\circ$, for which
$\chi^2_r = 5.2$ (for 13 degrees of freedom); the corresponding 15 systems are indicated as
blue data points in Figures~\ref{fig_dmz} and \ref{fig_aitoff}.
The fit to these data yields $H_n = 1830^{+120}_{-250}$~pc
and $n_0 = 0.014\pm0.001$~cm$^{-3}$, shown by the solid line in
Figure~\ref{fig_dmz}.
The bootstrap approach incorporates the added
uncertainties due to systematic deviations from the smooth underlying model.
Thus although $\chi^2_r \gg 1$, the errors in $H_n$ and $n_0$
meaningfully characterise the confidence intervals on
these parameters.

Our estimate of $H_n$ is substantially larger than that proposed
by previous authors, as summarised in Table~\ref{tab_n0}.  However,
our fit\footnote{We note that the best-fit parameters do not change
appreciably for larger values of $|b|_{\rm min}$: for example, for $|b|_{\rm
min} = 65^\circ$, the fitted parameters are $H_n = 1870^{+80}_{-280}$~pc
and $n_0 = 0.0139\pm0.001$~cm$^{-3}$, with $\chi^2_r = 6.0$.} is
clearly a good description of virtually all the high latitude data,
especially those pulsars at high $z$.  The implied integrated
electron column density toward the Galactic poles is DM$_0 \equiv
n_0 H_n = 25.97^{+0.02}_{-2.35}$~pc~cm$^{-3}$, consistent with the
upper limits from pulsars in the Magellanic Clouds, and 
comparable to the integrated vertical DM found in most previous studies.

The reason for our much larger estimate of $H_n$ is clear from
Figure~\ref{fig_dmz}: the high-$z$ pulsars and Magellanic Cloud
data together constrain DM$_0$ for any model.  Curves with high
values of $n_0$ must thus turn over early to match this required
value for DM$_0$, while those with low $n_0$ can extend to high
values of $z$ before saturating.  When fitting all data (shown by
the dashed line), the group of red and orange points at 300~pc~$\la
z \la 1000$~pc forces the best fit to have a comparatively high
value of $n_0$, and hence a low value for $H_n$.  The blue data
points all systematically lie below the other measurements, and the
corresponding fit (shown by the solid line) must then have a large
value of $H_n$ to converge to the required value for DM$_0$.

\subsection{Scale Height of Emission Measure}
\label{sec_em}

In \S\ref{sec_distrib}, we focused exclusively on pulsar DMs.
However, an independent measure of the free electron column along
the line of sight comes from measurements of the H$\alpha$ surface
brightness, which can then be used to infer the corresponding EM,
\begin{equation}
{\rm EM} = \int_0^{\infty} n^2(l) dl .
\label{eqn_em}
\end{equation}
The integral in Equation~(\ref{eqn_em}) differs from that for pulsar
DMs in Equation~(\ref{eqn_dm}) in that the former does not terminate
at any specific distance, and that the H$\alpha$ emission must be
corrected for dust extinction to obtain the corresponding EM. Thus
EMs and DMs along the same sightlines can only be meaningfully
compared for sources for which both $|b|$ and $z$ are simultaneously
high, so that extinction is low and the sources are entirely above
the WIM layer.\footnote{As discussed in \S\ref{sec_bmm06}, BMM06
and \cite{bm08}
attempted to correct for these effects for sources at
lower $|b|$ and $z$.} This means that EMs toward pulsars can be
used to infer volume filling fractions of ionised gas \citep[see][]{rey91},
but cannot be used to independently estimate the gas scale height,
as was done for DMs in Figure~\ref{fig_dmz}.

A separate approach to the EM scale height has been enabled by the
wide-field imaging spectroscopy of the WIM 
by the Wisconsin H$\alpha$ Mapper (WHAM). \cite{hrt99} used WHAM
to map the intensity of diffuse H$\alpha$ (i.e., that
emission not associated with individual \HII\ regions) for the restricted velocity
range corresponding to the Perseus spiral arm
as a function of $b$, and found
that this intensity dropped exponentially by a factor of 100 between
$b=-15^\circ$ and $b=-35^\circ$.  For a planar distribution of
ionised gas with an exponential density profile:
\begin{equation}
n^2(z) = (n^2)_0~e^{-\frac{z}{H_{n^2}}} , 
\label{eqn_n2z}
\end{equation} 
and assuming a distance to the Perseus arm of 2.5~kpc, \cite{hrt99}
derived a scale height for the EM layer of $H_{n^2} = 500 \pm
40$~pc. The distance to the Perseus arm has subsequently been
determined by trigonometric parallax to be 2.0~kpc \citep{xrzm06},
so we revise the published value of the EM scale height in Perseus
to $H_{n^2} = 400 \pm 30$~pc.

\subsection{The Filling Factor of Ionised Gas}
\label{sec_fill}

We have above derived two independent estimates of the WIM scale height:
$H_n = 1830^{+120}_{-250}$~pc using the dispersion measures of
pulsars at known distances in \S\ref{sec_high}, compared to $H_{n^2}
 = 400 \pm 30$~pc from the latitude-distribution of H$\alpha$ emission
in and above the Perseus arm, as discussed in \S\ref{sec_em}.

Comparison of Equations~(\ref{eqn_nz}) and (\ref{eqn_n2z}) clearly
requires that $H_n/H_{n^2} = 2$, which is in strong
disagreement with the observed value $H_n/H_{n^2} =
4.7_{-0.8}^{+0.4}$. As discussed
in \S\ref{sec_intro}, such discrepancies
can be resolved by introducing the
volume filling factor, $f$, such that along a given sightline, a
fraction $f$ of the line of sight intersects clouds of uniform
electron density $N$, with the rest of the path not contributing
to the observed EM or DM (e.g., because it is occupied by neutral
gas).\footnote{We have defined a line-of-sight filling factor, which
is not necessarily equivalent to the three-dimensional volume
filling factor. However, for reasonable assumptions about a clumpy
distributed medium, the two different filling factors are equivalent
(see \S2.1 of BMM06).}  In some local region, we then have:
\begin{equation}
n(z) \equiv f(z)~N(z),
\label{eqn_n}
\end{equation}
where $N$ is the electron density within a WIM cloud, and $n$ is the
mean electron density, averaged over some scale
significantly larger than the average cloud separation.
As originally proposed by \cite{kh87,kh88b} and discussed extensively
by BMM06, we can now separately consider
independent distributions of $f$ and $N$, both as a function of~$z$:
 \begin{eqnarray}
f(z) = f_0~e^{+\frac{z}{H_f}} ,
\label{eqn_fz} \\
N(z) = N_0~e^{-\frac{z}{H_N}} , 
\label{eqn_Nz} 
\end{eqnarray}
where $f_0$ and $N_0$ are the extrapolated filling factor and cloud
density at mid-plane, respectively, and $H_f$ and $H_N$ are the
corresponding scale heights of the two distributions.
Note that $f$ is an increasing exponential function of $z$ in
Equation~(\ref{eqn_fz}), since we expect the WIM to increasingly
dominate over neutral gas at progressively higher values of $z$
\citep{kh87,avi00}.

Substituting Equations~(\ref{eqn_fz}) and (\ref{eqn_Nz}) into 
Equation~(\ref{eqn_n}), we then obtain:
\begin{equation}
n(z) = f_0~N_0 e^{-\left(\left[\frac{1}{H_N} -
\frac{1}{H_f}\right]z\right)},
\end{equation}
so that from Equation~(\ref{eqn_nz}) we then infer:
\begin{eqnarray}
n_0 = f_0~N_0, 
\label{eqn_n0} \\
\frac{1}{H_n} = \frac{1}{H_N} - \frac{1}{H_f}.
\label{eqn_hdm}
\end{eqnarray}
Similarly, for density squared, we find:
\begin{equation}
n^2(z) = f_0~N_0^2 e^{-\left(\left[\frac{2}{H_N} -
\frac{1}{H_f}\right]z\right)},
\end{equation}
which implies:
\begin{eqnarray}
(n^2)_0 = f_0~N^2_0, 
\label{eqn_n2} \\
\frac{1}{H_{n^2}} = \frac{2}{H_N} - \frac{1}{H_f}.
\label{eqn_hem}
\end{eqnarray}

Equations~(\ref{eqn_hdm}) and (\ref{eqn_hem}) can then be solved jointly
to yield:
\begin{equation}
H_N = \frac{H_{n^2}~H_n}{H_n - H_{n^2}},
H_f = \frac{H_{n^2}~H_n}{H_n - 2H_{n^2}}. 
\end{equation}
Our adopted values
$H_n = 1830^{+120}_{-250}$~pc 
and $H_{n^2} = 400 \pm 30$~pc then imply separate scale heights.
\begin{eqnarray}
H_N = 510_{-40}^{+45}~{\rm pc}, 
\label{eqn_hn} \\
H_f = 700_{-70}^{+100}~{\rm pc}.
\label{eqn_hf} 
\end{eqnarray}
Note that despite the broadly comparable values of $H_N$ and $H_f$,
the error bars on these quantities are not gaussian,
and we can exclude the possibility that $H_N = H_f$ at high confidence. 

We can separately determine the normalisation factors $f_0$ and
$N_0$, as defined in Equations~(\ref{eqn_fz}) and (\ref{eqn_Nz}),
respectively, as follows. We determined in \S\ref{sec_high} that
pulsar data imply an extrapolated\footnote{Note that the fitted 
value of $n_0$ is not the true
mid-plane density, but that of the thick-disk component alone,
extrapolated to $z = 0$; see discussion in \S\ref{sec_mid}.}
mid-plane mean electron density
$n_0 = 0.014\pm0.001$~cm$^{-3}$.  The mean vertical component of
the H$\alpha$ intensity at high latitude, assuming a planar geometry
for the emitting region, is 0.68~rayleighs \citep{hrt+03}. For
electron temperatures in the range 8000 to 12\,000~K \citep{rht99,hrt99},
this corresponds to a total vertical emission measure EM$_0 \approx
1.9\pm0.3$~pc~cm$^{-6}$ \citep[the correction for dust extinction
at high latitudes is at the level of $\approx2\%$ and can be
ignored;][]{sfd98}.  If we combine Equations~(\ref{eqn_em}),
(\ref{eqn_n2z}) \& (\ref{eqn_n2}), we determine:
\begin{equation}
{\rm EM}_0 = f_0~N_0^2~H_{n^2}.
\label{eqn_em0}
\end{equation}
Assuming that the observed H$\alpha$ emission at high latitudes
originates from the diffuse WIM \citep[see][]{hbk+08},
we can combine Equations~(\ref{eqn_n0}) \& (\ref{eqn_em0}) to yield:
\begin{eqnarray}
N_0 = \frac{f_0 N_0^2}{f_0 N_0} = \frac{{\rm EM}_0}{H_{n^2}~n_0} 
= 0.34\pm0.06~{\rm cm}^{-3}, 
\label{eqn_N0} \\
f_0 = \frac{(f_0 N_0)^2}{f_0 N_0^2} = \frac{(n_0)^2~H_{n^2}}{{\rm EM}_0} 
= 0.04\pm0.01 .
\label{eqn_f0}
\end{eqnarray}
The calculated values of $N_0$ and $f_0$ both clearly have a linear
dependence on the assumed value of EM$_0$.  The estimates for $N_0$
and $f_0$ will correspondingly change if a slightly different value of
EM$_0$ is assumed \citep[e.g.,][]{hbk+08}, but the values for all other
quantities discussed above (i.e., $n_0$, DM$_0$, $H_n$, $H_{n^2}$, $H_N$
\& $H_f$) will be unchanged.

For the sake of completeness and for comparison with previous studies,
we also compute the vertical ``occupation length'', $L_c \equiv
({\rm DM}_0)^2/{\rm EM}_0$, and ``characteristic density'', $N_c \equiv
{\rm EM}_0/{\rm DM}_0$, of the WIM \citep{rey91,hbk+08}.
For DM$_0 \approx 26.0$~pc~cm$^{-3}$ and
EM$_0 \approx 1.9$~pc~cm$^{-6}$ as derived above, we find $L_c \approx
350$~pc and $N_c \approx 0.07$~cm$^{-3}$. However, we caution that
in the case where $H_n / H_{n^2} \neq 2$ as argued here, these
quantities have no direct physical interpretation --- see further
discussion in \S\ref{sec_mid}.

\section{Discussion}
\label{sec_disc}

Our main result is that within a few kpc of the Sun, the scale height
of the thick disk of warm ionised gas in the Galaxy is $H_n \approx1.8$~kpc. This is
a factor of $\sim2$ higher than the value that has been 
found in most previous studies (see summary in Table~\ref{tab_n0}).
If this is a global property of the
Galactic ISM, this has a number of important implications.

However, it is important to note that our large value for $H_n$
results from our exclusion of low-latitude pulsars, whose DMs require
a higher value of $n_0$, and hence a lower value of $H_n$,
for constant DM$_0$. Because our sample consists only of high-$|b|$
pulsars, it could be argued that we are probing the WIM immediately
above and below the Sun, and thus may be tracing local structure
rather than any overall properties of the Milky Way
\cite[see][]{rey77,rey91b}.

In the following discussion, we first consider the possible presence
of local structures in DM and in EM, and conclude that our results
are not biased by any such features. Next, we examine our implicit
underlying assumption that a single phase of the WIM simultaneously
produces both DMs and EMs, and confirm that ths assumption is
justified by the data.  In the rest of this section, we then consider
some more general implications for the properties of warm ionised
gas at increasing distances from the Galactic mid-plane.

\subsection{The Effects of Local Structure}
\label{sec_struc}

\subsubsection{Local Structure in Dispersion Measure}
\label{sec_struc_dm}

In Figure~\ref{fig_overhead} we show the locations of the pulsars in
our sample projected onto the Galactic plane, overlaid with the NE2001
electron density model.  Of the 15 systems for which $|b| \ge 40^\circ$
(shown in blue), six sight lines (all to globular clusters) extend over
a significant distance through the Galaxy, traversing several, separate,
spiral arms.  These data clearly do not probe merely local gas, but are
representative of the WIM within several kpc of the Sun. In any case,
these sightlines should not be dominated by individual density features,
since the long path lengths ensure that individual clumps
or cavities in the electron distribution largely average out.

The remaining nine systems (all pulsar parallaxes) lie much closer
to the Sun. As indicated in Figure~\ref{fig_overhead}, there are
several cavities and enhancements in the solar neighbourhood
\citep[see ][and references therein]{cr87,tbm+99,slcw99}, whose
presence could dominate the DM of nearby pulsars, and thus might
bias our fits.  A qualitative argument against this possibility is
that in Figure~\ref{fig_dmz}, the envelope defined by the high-$|b|$
parallax data joins smoothly onto that traced out by the high-$|b|$
globular cluster data.  Since we have argued above that the DMs to
globular clusters trace the widespread properties of the WIM in the
disk, the pulsar parallaxes are also consistent with this interpretation.

We can quantify the effects that local structure might be having
on the data by plotting ${\rm DM}/d$ versus $z$, where $d$ is the
distance to each source.  This distribution is plotted in
Figure~\ref{fig_nz}: the electron column per unit distance plotted
in the ordinate is equal to the mean electron density along the
sightline to a given source.  The corresponding predictions for
${\rm DM}(z)/d \equiv {\rm DM}_\perp(z)/z$ for each of the curves
drawn in Figure~\ref{fig_dmz} are plotted in  Figure~\ref{fig_nz},
with the form of ${\rm DM}_\perp(z)$ given by Equation~(\ref{eqn_dmz}).

For a generally smooth distribution of electrons in all directions,
values of ${\rm DM}/d$ will follow a single curve such as those
plotted in Figure~\ref{fig_nz}.  Pulsars located inside or just
beyond a cavity will exhibit values of ${\rm DM}/d$ that sit below
this curve, while pulsars that sit behind density enhancements will
have values of ${\rm DM}/d$ above this curve. Both such effects are
clearly seen in Figure~\ref{fig_nz}, especially for sources at low
$|b|$ (coloured red and orange), which are not included in our best
fit.  However, if local structure is affecting our data, then
high-latitude pulsars (coloured blue) at low values of $z$ should
have values of ${\rm DM}/d$ that deviate more from the best fit
curve than do data for pulsars at higher $z$.  Only for one pulsar
(at $z \approx 300$~pc) of the 15 sources in our
sample is there a slight suggestion of such an effect.  The small
errors in $n_0$ and $H_n$ derived from bootstrapping confirm that
these data do not have a significant effect on our overall fit.

From Figures~\ref{fig_overhead} and \ref{fig_nz}, we conclude that
localised structures are not biasing our results, and that the fits
to DM data in Figure~\ref{fig_dmz} are a meaningful representation
of the ensemble scale height of diffuse ionised gas within several
kpc of the Sun.

\subsubsection{Local Structure in Emission Measure}
\label{sec_struc_em}

Another way in which localised structure may bias our measurements
is in our assumption that the scale heights $H_{n^2}$  and $H_n$
correspond to the same regions.  The determination of $H_{n^2}$
by \cite{hrt99} was specific to a particular segment of the
Perseus arm. In contrast, the DM data in Figure~\ref{fig_dmz}
represent integrated column densities spread over a factor of
$\approx200$ in $z$ and distributed over a range of Galactic
longitudes (see Figs.~\ref{fig_aitoff} and \ref{fig_overhead}).
Thus while we have argued in \S\ref{sec_struc_dm} that the DM data
represent ensemble properties over several kpc of the Galactic disk,
the EM data provide no such guarantees.  If ionised gas above the
Perseus arm has special properties (and indeed Fig.\ \ref{fig_overhead}
shows that none of our high-latitude pulsars lie in or behind the
Perseus arm), then the conclusions of \S\ref{sec_fill} are not
correct.

There are two arguments that support the calculations in \S\ref{sec_fill}.
First, Lazio \& Cordes (1998a,b\nocite{lc98b,lc98c}) have specifically
studied the properties of diffuse ionised gas toward the Perseus
arm and over the rest of the outer Galaxy, and conclude that the
ionised disk shows neither warping nor flaring in these regions.
Second, even if structure peculiar to the Perseus arm might introduce
some bias into our calculations, the fundamental result in
\S\ref{sec_fill} is that $H_n / H_{n^2} \gg 2$. Only completely
different behaviour for the Perseus arm compared to the sightlines
in other directions probed by pulsar DMs could reduce this ratio
sufficiently to remove the need for a $z$-dependent filling factor.
NGC~891 and M31, two spiral galaxies with broadly similar properties
similar to the Milky Way, show variations in the thickness of their
ionised gas layers as a function of position, but at fractions of no
more than $\sim$50\% \citep{rkh90,mv03,fbbs04}.

\subsection{Two Cloud Populations?}
\label{sec_sep}

In \S\ref{sec_fill}, we assumed that at a given height $z$, WIM
clouds can be described by a single volume-averaged electron density,
$n(z)$.  This is obviously an idealised situation, but is a reasonable
approximation of actual data, which suggest that $n(z)$ shows a
modest spread in density around a well-defined average value
\citep{hbk+08}.  In the model considered in \S\ref{sec_fill}, we
have assumed that this simple cloud population produces both the DMs
and EMs that we observe.  The
fact that $H_n / H_{n^2} \ne 2$ can then be explained by invoking
separate dependences of $N$ and $f$ on $z$, as described by
Equations~(\ref{eqn_fz})~\&~(\ref{eqn_Nz}).

However, an alternative possibility is that there are (at least)
two populations of WIM clouds, with separate densities, filling
factors and scale heights. Specifically, if we suppose that one set
of clouds has a low density and a high filling factor, while a
second cloud population have a much higher density and a much lower
filling factor, then observations of DM will only probe the first
group, while data on EM will only trace the second \citep{gou71,hei01}.
Carrying out a joint analysis of DMs and EMs to infer overall
properties of $N$ and $f$, as we have done in \S\ref{sec_sep}, would
then not be meaningful.

We thus now consider whether the data can be validly interpreted
in terms of two separate populations of
WIM clouds.  Suppose that one group
of WIM clouds has electron density $N_A(z)$, filling factor $f_A$ 
(which is not a function of $z$)
and volume-averaged density $n_A(z) \equiv f_A N_A(z)$. The density of these
clouds is distributed exponentially, such that $n_A(z)$ and $N_A(z)$
both have a scale height $H_A$.
A second cloud population has corresponding parameters $N_B(z)$,
$f_B$ and $n_B(z)\equiv f_B N_B(z)$, with an exponential scale
height $H_B$.

The vertical dispersion and emission measures integrated through
the thick disk then become:
\begin{eqnarray}
{\rm DM}_0 = f_A N_{A,0} H_A + f_B N_{B,0} H_B~, \\
{\rm EM}_0 = f_A N^2_{A,0} H_A/2 + f_B N^2_{B,0} H_B/2~, 
\end{eqnarray}
respectively, where the subscript ``0'' indicates extrapolated
mid-plane values, and where we have disregarded second order terms
in $f_A f_B$. The observed behaviours of DM and EM with $z$ are both well
described by a single exponential layer (Fig.~\ref{fig_dmz}; Fig.~4 of
Haffner \etal\ 1999\nocite{hrt99}), implying that only a single population
of clouds contributes to each set of measurements. We thus adopt
the requirement that the DMs only probe population~A 
but that the EMs only probe population~B:
\begin{eqnarray}
f_B N_{B,0} H_B \ll {\rm DM}_0~, \label{eqn_fb} 
\end{eqnarray}
and
\begin{eqnarray}
f_B N^2_{B,0} H_B/2 \approx {\rm EM}_0 . \label{eqn_fa}
\end{eqnarray}
Comparing Equations~(\ref{eqn_em0}) \& (\ref{eqn_fa}), we can then
identify $H_B/2 \equiv H_{n^2}$, so that combining Equations~(\ref{eqn_fb})
\& (\ref{eqn_fa}) then yields:\footnote{A similar analysis for
population~A implies $f_A \gg 0.1$ and $N_{A,0} \ll 0.15$~cm$^{-3}$.}
\begin{eqnarray}
f_B \ll \frac{L_c}{4 H_{n^2}} \approx 0.2 , \\
N_{B,0}  \gg 2 N_c \approx 0.15~{\rm cm}^{-3},
\end{eqnarray}
with $L_c$ and $N_c$ defined at the end of \S\ref{sec_fill}.
We can rule out the existence of such a component to the diffuse
WIM, because H$\alpha$ data lead to direct estimates of mid-plane
densities for the diffuse WIM of $N_{B,0} \approx 0.1-0.3$~cm$^{-3}$
\citep{rey80,rey83,rtk+95}. Since the predicted cloud densities of
population~B are at odds with those observed, we conclude that it
is not valid to argue that DMs and EMs are produced by separate
populations. The alternative is that the distribution of ionised
clouds is described by a single set of parameters, $n(z) \equiv f(z) N(z)$,
as laid out in \S\ref{sec_fill}.

\subsection{General Implications}
\label{sec_gen}

We have shown in \S\ref{sec_struc} that our results are not likely
to be merely a result of local structure in DM or EM, and we have
demonstrated in \S\ref{sec_sep} that we have not mistakenly conflated
data on two separate populations of ionised clouds.  We emphasise
that even if some of our arguments in these Sections can be shown
to have shortcomings, our main result, that $H_n  \approx 1.8$~kpc,
is derived purely from pulsar DMs, and does not rely on any subsequent
assumptions about filling factors or EMs.  We now consider some of
the implications of our revised WIM parameters.

\begin{figure*}
\centerline{\psfig{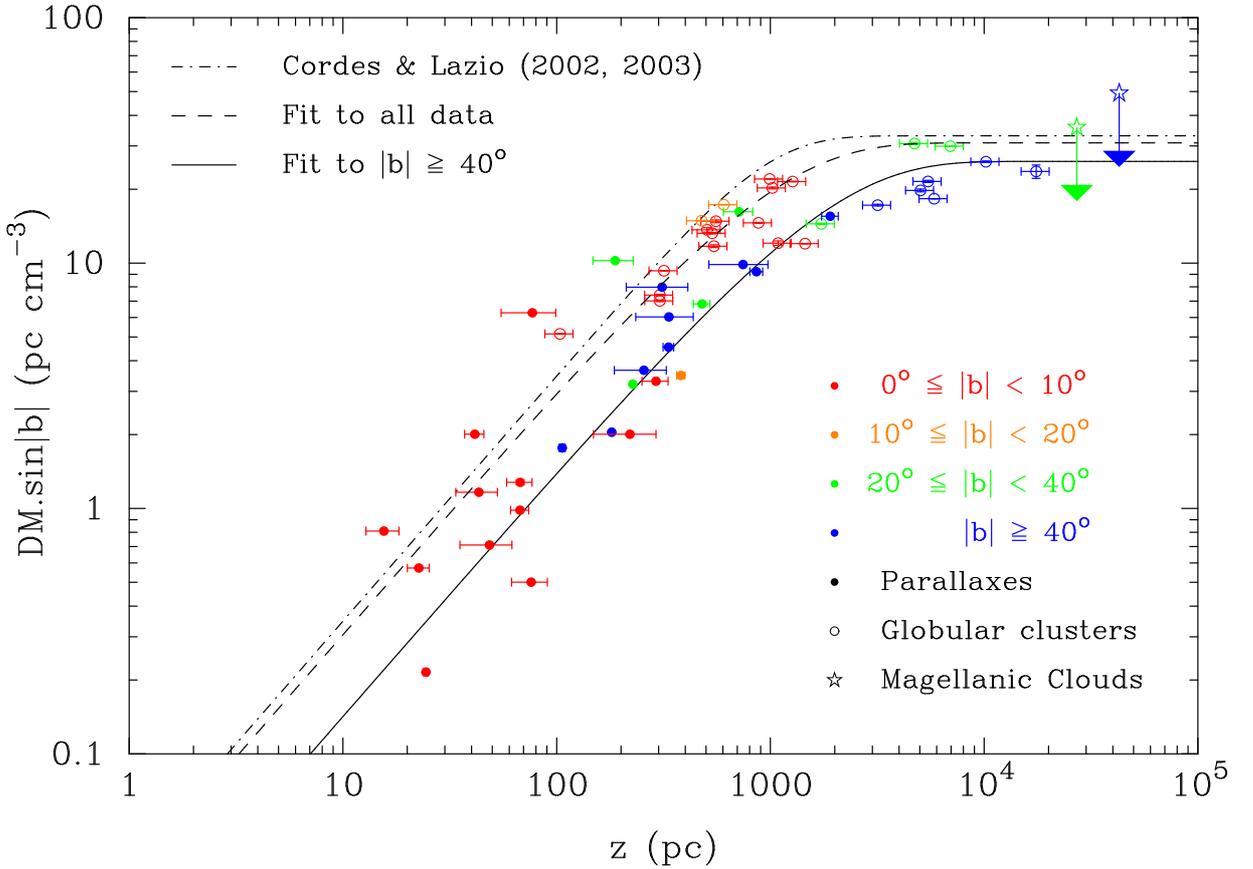}}
\caption{Distribution of dispersion measure as a function of height
above the Galactic plane for pulsars with known, reliable, distances.
Pulsars with distance determinations from trigonometric parallaxes,
from associations with globular clusters and from associations with
the Magellanic Clouds are shown as filled circles, open circles and
stars, respectively (data for the Magellanic Clouds are indicated
as upper limits, since the Galactic and Magellanic DM contributions
cannot be separated). Data are colour-coded according to the magnitude
of their Galactic latitude, $|b|$, as indicated in the legend. The
error bars represent the published uncertainties. The three curves
represent planar slabs of ionised gas with a scale
height $H_n$ for the volume averaged electron density, $n$. 
The dot-dashed line shows a $n \propto
{\rm sech}^2 (z/H_n)$ density dependence, using the parameters given
in Table~3 of \cite{cl02} (see also Fig.~1 of \citealt{cl03}); the
dashed line shows the best fit for an exponential slab ($n \propto
e^{-z/H_n}$) to all the data in the figure; the solid line
shows the best exponential fit to only those data for which $|b| \ge 40^\circ.$}
\label{fig_dmz} 
\end{figure*}

\begin{figure*}
\centerline{\psfig{file=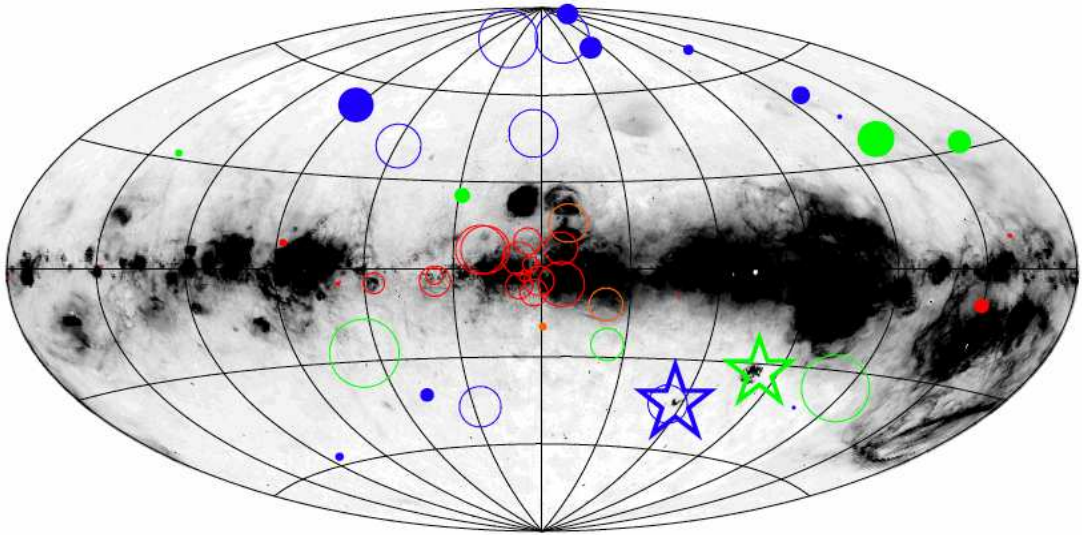,width=\textwidth}}
\caption{Aitoff projection in Galactic coordinates of pulsars with
known, reliable, distances, super-imposed on the all-sky distribution
of H$\alpha$ intensity \citep[as derived by][]{fin03}.  
The Galactic Centre is at the centre
of the projection; colours and symbols are as in Figure~\ref{fig_dmz}.
The centre of each symbol marks the coordinates of the corresponding
source, while the diameter of each symbol is proportional to
$\log_{10}({\rm DM}.\sin |b|)$ for that source.
The greyscale is linear, ranging from 0 to 20 rayleighs.}
\label{fig_aitoff}
\end{figure*}

We first consider the Galactic distribution of pulsars. The most
frequent application of models for the WIM is to derive distance
estimates of pulsars from measurements of their DMs.  Figure~\ref{fig_dmz}
demonstrates that the NE2001 model does not correctly reproduce the
DMs of high-latitude pulsars at known distances. In turn, it seems
likely that the DM-derived distances for many high-$|b|$ pulsars without
independent distance constraints may be substantial underestimates;
the corresponding predictions from NE2001 should be used with
caution. Indeed, \cite{kbm+03} and \cite{lfl+06} found that if
distances predicted by NE2001 are applied to the Parkes
multibeam pulsars, these sources are then anomalously clustered toward small $z$.

A revision of the distances to high-latitude pulsars would have
important implications. Such data play a major role in characterising
the population statistics of millisecond (which tend to be found
at high $|b|$ due to their large ages), as well as to predict the
yields of future pulsar surveys.  Recent studies of these issues
have all been underpinned by the NE2001 model
\citep[e.g.,][]{hllk05,sgh07}, and the corresponding conclusions
may now need to be reconsidered. Specifically, if the true distances
to high-$|b|$ pulsars are larger than have been assumed in recent
studies, then millisecond pulsars may have higher space velocities,
be more luminous, and be more numerous than has been previously
argued.

Overall, we note that
while the NE2001 electron density model most likely underestimates
the distances to {\em high}-latitude pulsars, one of the major
strengths of this model was that it corrected the severe over-estimates
that previous models had made for the distances to {\em low}-latitude
pulsars. Thus it is unwarranted to argue that the entire NE2001
model is unreliable; our analysis focuses only on the high-latitude
component, and indicates only that this aspect needs revisiting.
Correspondingly, we do not recommend that Equation~(\ref{eqn_dmz})
or Figure~\ref{fig_dmz} be used 
to estimate the distances or DMs to individual pulsars at high
$|b|$, but instead suggest that in future
studies, the results derived here should be incorporated into a revised
global
model for the disk, spiral arms and halo.

As noted in \S\ref{sec_ne2001}, a full three-dimensional model of
Galactic magnetic fields, cosmic rays and thermal electrons, but
assuming a warm layer with scale height $H_n \sim1$~kpc, results
in an unacceptably high halo magnetic field strength of $\approx10$~$\mu$G,
and an unreasonable truncation of cosmic ray electrons at $z=1$~kpc
\citep{srwe08}.  As \cite{srwe08} explicitly discuss, a WIM layer
with a scale height of $\sim2$~kpc can simultaneously solve both
these problems. The independent conclusions that we have arrived
at here thus support a relatively weak ($\sim2$~$\mu$G) magnetic field in the halo,
and a cosmic ray distribution which can diffuse to high $z$.

\begin{figure*}
\centerline{\psfig{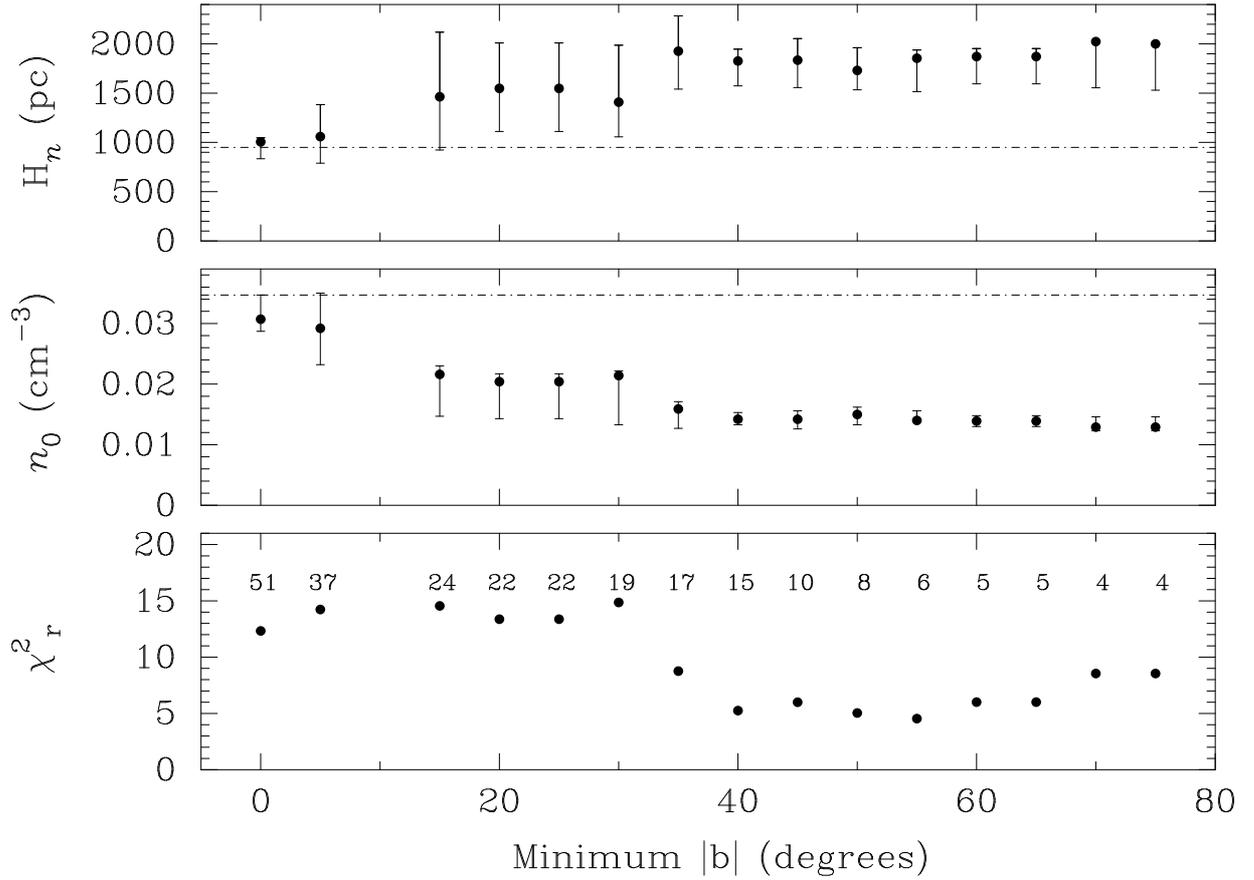}}
\caption{
Best fit parameters for a planar slab of ionised gas with an
exponential scale height, as a function of the minimum cut-off used
for $|b|$. The top, centre and bottom panels show the scale height,
$H_n$, the extrapolated mid-plane electron density, $n_0$, and
the reduced $\chi^2$ of the fit, $\chi^2_r$, respectively. In the
top two panels, the dot-dashed line shows the parameters given in
Table~3 of \cite{cl02}, while the error bars show the 68\% confidence
intervals on each parameter. In the bottom panel, the number above
each datum indicates the number of points from Figure~\ref{fig_dmz}
used in the fit. The fit did not converge for a cut-off $|b|_{\rm min} =
10^\circ$.}
\label{fig_data} 
\end{figure*}

\begin{figure*}
\centerline{\psfig{file=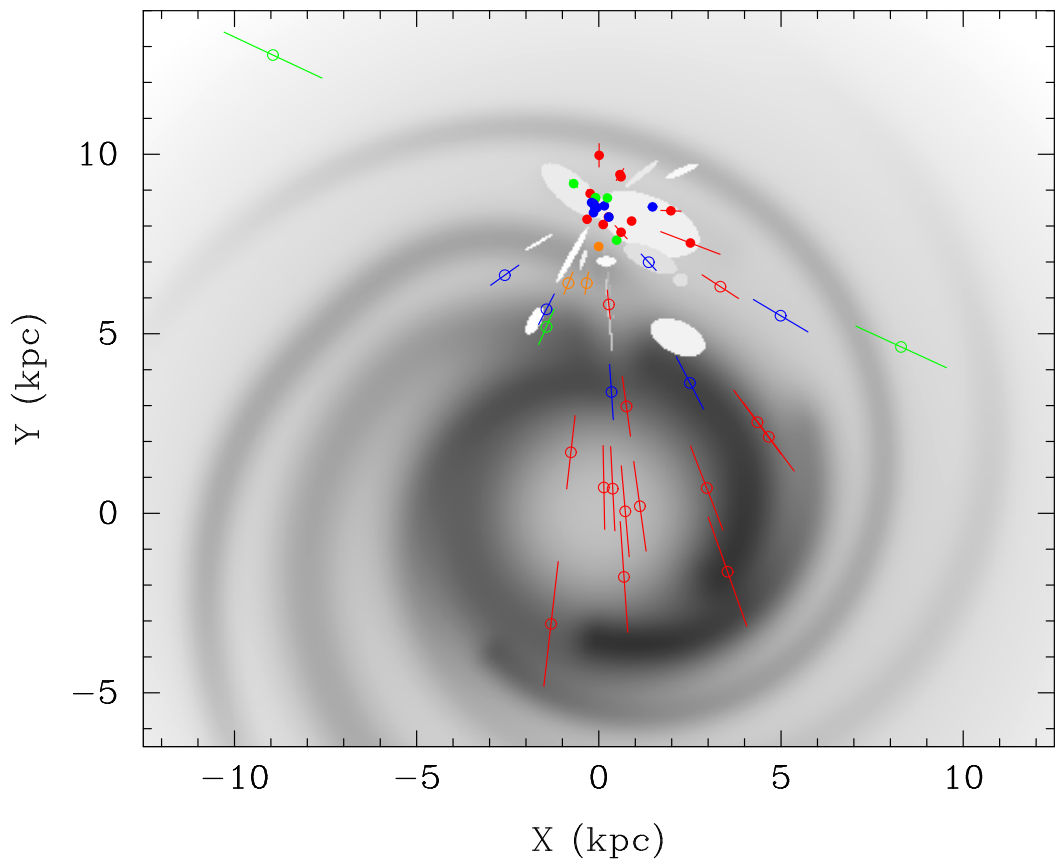,width=\textwidth,angle=270}}
\caption{The Milky Way as viewed from the North Galactic Pole, with
the Galactic Centre located at the origin of the coordinate system
and the Sun at $(0~{\rm kpc},~8.5~{\rm kpc})$.  The greyscale
represents the distribution of mid-plane electron density from the
NE2001 model of \cite{cl02} (for clarity, individual high-density
clumps are not shown).  The data show the projected locations onto
the Galactic $(X,Y)$ plane of pulsars with known, reliable, distances.
Colours and symbols are as in Figure~\ref{fig_dmz}.}
\label{fig_overhead}
\end{figure*}

\begin{figure*}
\centerline{\psfig{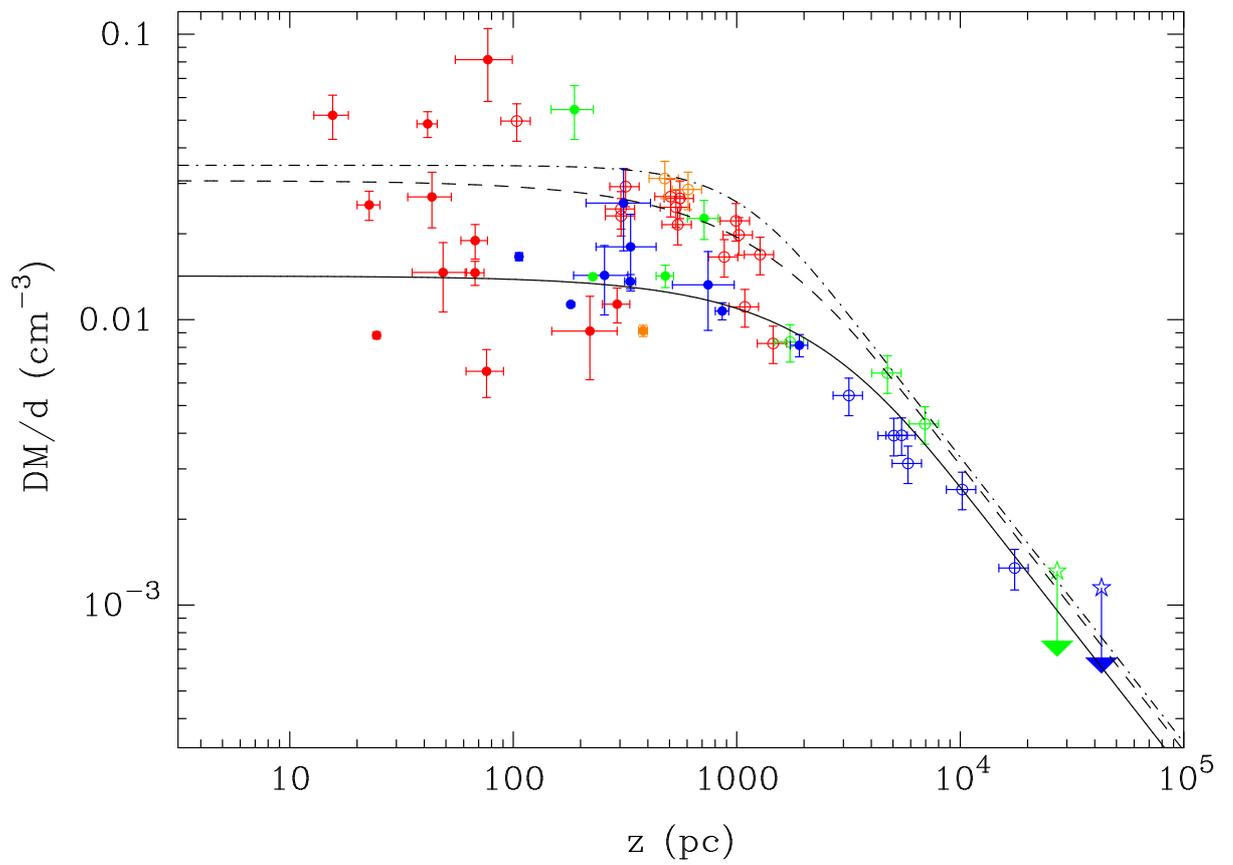}}
\caption{Dispersion measure per unit distance (equivalent to the
electron density averaged along the line of sight) as a function
of height above the Galactic plane for pulsars with
known, reliable, distances.  Colours, symbols and curves are as in
Figure~\ref{fig_dmz}.}
\label{fig_nz}
\end{figure*}

Despite the more consistent overall picture of the WIM that our
analysis can potentially provide, we note that even after artificially
increasing the errors in $z$ as was carried out in \S\ref{sec_all},
the best fit in Figure~\ref{fig_data} has $\chi^2_r \approx 5$. The
fact that $\chi^2_r \gg 1$ clearly indicates that a smooth exponential
slab is not an accurate description of the individual data points.
As discussed extensively by Cordes \& Lazio (2002, 2003\nocite{cl02,cl03}),
many sightlines through the Galaxies pass through intervening regions
where the density is either enhanced (clumps) or reduced (voids).
These systematic deviations are not incorporated in our model, and
imply that $\chi^2_r$ will always be relatively large in any simple
fit to this complex data set. However, it is noteworthy that
$\chi^2_r$ does drop significantly for higher cut-offs in $|b|$
(see Fig.~\ref{fig_data}), indicating that the WIM 
becomes more homogeneous as one moves further from the Galactic
plane. We further note in Figure~\ref{fig_dmz} that the low-latitude
data (in red) are distributed both above and below the fits that
include these data (dashed and dot-dashed lines),
while the highest latitude data (in blue) generally lie only
on or above the corresponding best fit (solid line). 
This suggests that in the Galactic plane
there are both clumps and voids in the WIM, but that far from the
plane any deviations from a smooth distribution are mainly in the
form of clumps, rather than voids. This same argument produces
additional qualitative evidence against the NE2001 model for ionised
gas at high $z$. If the NE2001 description (shown as a dot-dashed
line in Fig.~\ref{fig_dmz}) is correct, then the fact that almost
every blue data-point in Figure~\ref{fig_dmz} lies well below this
line requires that there be separate voids in electron density in
front of all these high-$|b|$ pulsars. A large number of voids to
account for all these low DMs does not seem warranted, especially
given that these pulsars are distributed at a wide range of distances,
in a variety of directions, and sit both above and below the Galactic
plane (see Figs.~\ref{fig_aitoff} \& \ref{fig_overhead}).  The
sustained low DMs for all these pulsars compared to the NE2001 model
suggests that a different overall density distribution is required
in these directions.

\subsection{Mid-Plane Properties of the Diffuse WIM}
\label{sec_mid}

In this section, we compare our extrapolated mid-plane values for
the thick-disk WIM to other studies of the ISM in the disk.  
At $z=0$, we infer a volume-averaged density for the diffuse WIM
of $n_0 = 0.014\pm0.001$~cm$^{-3}$.
Observationally, this can be compared to values of ${\rm
DM}/d$ for pulsars at low $|b|$ with known distances, which typically
yield $n_0 \sim 0.03-0.1$~cm$^{-3}$ along complicated sightlines
through spiral arms, and $n_0 \approx 0.01-0.02$~cm$^{-3}$ in
inter-arm regions \cite[][and references therein]{jkww01,wsx+08}.
As discussed in \S\ref{sec_intro}, this indicates two components
for warm ionised gas in the ISM: free electrons associated with
individual \HII\ regions (only found in spiral arms at low $|b|$),
and the diffuse, thick-disk WIM (seen at all latitudes and in all directions).
We have not included any contribution from the former component in
our model, since we only fit DMs at high $|b|$ and high $z$.  Thus
the fact that observational estimates for $n_0 = {\rm DM}/d$ in
inter-arm regions agree well with our derived value for $n_0$
supports the picture of an extended diffuse WIM in which individual
dense ionised regions are embedded, and suggests that the thick WIM
layer extends all the way down into the Galactic plane.

Estimates of $n_0$ from previous attempts to globally model the thick-disk WIM are
summarised in Table~\ref{tab_n0}.  Our value for $n_0$ is systematically
lower than that in almost all these other studies.
In most cases, this results from the fact that
${\rm DM}_0 \equiv n_0 H_n$ for an exponential slab.  Since all studies
generally agree on ${\rm DM}_0$, but most previous authors have derived
much smaller scale heights than we have found, their estimates of $n_0$
are correspondingly larger.  BMM06 set only a lower limit on $H_n$; their
estimate of $n_0$ may be biased by the systematic distance uncertainties
in their analysis, as discussed in \S\ref{sec_bmm06}.

A variety of studies have also estimated the mid-plane filling factor
of the WIM, as summarised in Table~\ref{tab_f0}.  The value $f_0 =
0.04 \pm 0.01$ which we derived in \S\ref{sec_fill} agrees with the
estimate of BMM06, but is substantially smaller than that found in most
other studies. This discrepancy results from the fact that calculations
of $f_0$ are usually degenerate with estimates of $n_0$ (see, e.g.,
Eqn.~[9] of \citealt{rey91}). In our model, $f_0 \propto (n_0)^2$ (from
Eqn.~[\ref{eqn_f0}]), and we have explained above that $n_0$ needs to
be revised downward from previous estimates by a factor of $\sim1.7-2$.
The true value of $f_0$ in the WIM must then be a factor of $\sim3-4$
smaller than what has been found in previous studies, as we have 
determined here.

Our estimates for $f_0$ and $n_0$ imply a typical mid-plane density
for WIM clouds $N_0 \equiv n_0/f_0 = 0.34\pm0.06$~cm$^{-3}$, which
agrees with the theoretical prediction $N_0 = 0.3$~cm$^{-3}$ of
\cite{cox05}.  Independent estimates of $N_0$ have been carried out
by \cite{rey80,rey83} and \cite{rtk+95}, who measured EMs along a
variety of sightlines and found $N_0 \approx 0.1-0.3$~cm$^{-3}$,
in reasonable agreement with our value.  The thermal
pressure of the WIM is\footnote{The factor of 2 corresponds to a
fully ionised hydrogen gas.} $P = 2 N k T_e$, so for $T_e \approx
8000$~K \citep[e.g.,][]{rht99}, we infer a mid-plane ISM pressure
for the thick-disk WIM of $P_0/k \approx 5400\pm1000$~K~cm$^{-3}$.
Other estimates for the pressure of the WIM and of other ISM phases
show a wide variation depending on local conditions, but certainly
the above value falls comfortably within the typically observed
range \citep[see][]{cox05}.

\cite{hei01} argued that there must be two separate components of
the diffuse WIM, one which contributes to EMs seen in H$\alpha$,
and a second which separately produces the DMs seen for pulsars.
We have already shown in \S\ref{sec_sep}
that, as an overall model for the ISM, a two-phase WIM is
not supported by the data. Here we briefly consider the specific arguments
made by \cite{hei01}. He
begins by assuming that the EMs and DMs are produced
in the same ISM phase, and then writing $N_0 = N_c \equiv {\rm EM}_0/{\rm DM}_0$.
The implied density leads to an anomalously underpressured WIM
compared to the cold, neutral, ISM, a problem which can be relieved
by putting the electrons that contribute to EMs and DMs into distinct
phases.  However, the expression $N_0 = {\rm EM}_0/{\rm DM}_0$ is only valid
for a slab of uniform density; for the exponential fall-off considered
here, the appropriate expression is $N_0 = ( {\rm EM}_0 / H_{n^2})/ ({\rm DM}_0
/ H_n)$, which leads to a WIM pressure $\sim4-5$ times higher than
in the discussion of \cite{hei01}. With this correction, there is
approximate pressure equilibrium between the WIM and other phases
of the ISM, so that a model in which the EMs and DMs are produced
by the same gas is sustainable.

Finally, we can combine Equations~(\ref{eqn_n2}) \& (\ref{eqn_em0})
to estimate a mean square electron density at midplane $(n^2)_0 \equiv
{\rm EM}_0/H_{n^2} = 0.005\pm0.001$~cm$^{-6}$.  This value, estimated
from quantities which we have not substantially revised in this
paper, is broadly comparable to previous estimates, as listed
in Table~\ref{tab_n20}.

In summary, our results imply that within a few kpc of the Sun, the thick-disk
component of the WIM
at mid-plane is comprised of clouds of electron density $N_0 \approx
0.3$~cm$^{-3}$ and a volume filling factor of $f_0 \approx4\%$,
corresponding to a volume-averaged electron density $n_0 \approx
0.014$~cm$^{-3}$. If the diffuse WIM is considered separately from denser
gas in the plane associated with discrete \HII\ regions, these values are
in broad agreement with independent estimates made by other approaches.
We note that radio recombination line studies generally identify higher
values of $N_0$ and lower values of $f_0$ than what we have derived here
\citep[e.g.,][]{hrk96,ra01}, but these emission lines preferentially
trace denser gas, possibly associated with the outer envelopes of \HII\
regions or with the walls of chimneys in the ISM \citep{ana86,hrk96}.

\subsection{Anti-Correlation of $N$ and $f$, and Evolution of WIM
Properties with $z$}

The total mass and ionisation rate of the WIM are unchanged in our
analysis compared to
previous estimates \citep[e.g.,][]{rey90b}, since the calculations
of these quantities
primarily depend on the integrated column densities EM$_0$ and
DM$_0$, rather than on the scale-height.  What the calculations in
\S\ref{sec_fill} do imply is contrasting behaviours for the cloud
density, $N$, and the locally averaged electron density, $n$, as a
function of height above the Galactic plane. The former decays over
a relatively small scale height, $\approx500$~pc.  But the volume
filling factor, $f$, of these ionised clouds increases as one moves
further from the plane, counter-balancing the decreasing internal
cloud density.  Since $f$ increases with $z$ at a rate slightly
more slowly than that with which $N$ decays, $n \equiv fN$ decays
relatively slowly as a function of $z$.  These results agree with
the earlier findings of BMM06, who through a separate approach derived
broadly comparable values of $H_N$ and $H_f$ to those found here,
albeit with much larger uncertainties (compare Eqn.~[\ref{eqn_bmm2}]
with Eqns.~[\ref{eqn_hn}] \& [\ref{eqn_hf}]).

An important distinction between our analysis and that of BMM06 is
that we find $H_N < H_f$, while BMM06 concluded that $H_N \approx
H_f$. As per Equation~(\ref{eqn_hdm}), BMM06 thus could infer only a
lower limit on $H_n$, and indeed they argued that there is no
evolution in $n$ for $0 < z < 1$~kpc, based on the constancy of
${\rm DM}/d$ over this range.  However, as discussed in \S\ref{sec_bmm06},
there are large uncertainties in the values of $d$ and $z$ used in
their analysis.  Figure~\ref{fig_nz} shows that for our sub-sample
of pulsars with reliable distances and with $|b| \ge 40^\circ$,
there is a gradual decline in ${\rm DM}/d$ with increasing $z$
up to $z \approx 1$~kpc,
which
joins smoothly to the behaviour seen for $z \gg 1$~kpc (note that
the scatter for the blue points in Fig.~\ref{fig_nz} is much smaller
than for the equivalent plot in Fig.~12 of BMM06).  This can only
occur if $n$ steadily decays with $z$, requiring $H_N < H_f$ as we
have inferred above.

The decaying exponential for $N$ compared to the growing exponential
in $f$ obviously implies an overall anti-correlation between $f$
and $N$, as has also been found through various other approaches
\citep[][BMM06]{pyn93,ber98,bm08}.  For a dependence:
\begin{equation}
f(N) = \beta \left( \frac{N}{{\rm cm}^{-3}}\right)^{\alpha} , 
\label{eqn_fN}
\end{equation} 
we can combine Equations (\ref{eqn_fz}), (\ref{eqn_Nz}), (\ref{eqn_hn}),
(\ref{eqn_hf}), (\ref{eqn_N0}) \& (\ref{eqn_f0}) to derive $\alpha
= -0.73^{+0.06}_{-0.03}$ and $\beta = 0.018\pm0.004 $.  This determination
differs substantially from $\alpha = -1.07\pm0.03$ found by BMM06,
but agrees with $\alpha \approx -0.7$ as derived by \cite{pyn93}.
Since $\alpha \equiv -H_N/H_f$, the discrepancy between our estimate and
that of BMM06 simply results from our finding that $H_N < H_f$ as
discussed above.

BMM06 consider two possible physical interpretations of
Equation~(\ref{eqn_fN}) \citep[see also discussion by][]{bm08}:
either it can represent a stochastic, turbulent ensemble of clouds
of different sizes and different densities, or it describes the
equation of state of ionised clouds, in that it probes the response
of the WIM to a change in external conditions.  In the situation
that we consider here, the variation of $f$ and $N$ are both modeled
as smooth functions of $z$, so the latter case applies.

In this context,
what is the physical significance of the anti-correlation between
$N$ and $f$, and of the differing scale heights for these two
quantities?  We first note that the thermal gas pressure, $P = 2 N
k T_e$,  depends on $N$ but not on $f$. If the evolution of $T_e$
with $z$ is modest \citep{rht99,pw02}, then the pressure decays
exponentially with a scale height $H_N \sim 500$~pc, as described
by Equation~(\ref{eqn_Nz}).  This agrees well with the pressure scale height
derived from hydrostatic equilibrium calculations
of the full multi-phase ISM
\citep{bc90,fer98a}.

To interpret the evolution of $f$ with $z$, we consider the simple
case in which the WIM is composed of a large number of clouds, all
with the same number of electrons per cloud, $\mathcal{N}$, and
with $C dz$ clouds contained in a vertical interval
$dz$.  The volume of an individual cloud is $V(z)$, so that $N(z)
= \mathcal{N} / V(z)$.  If the total surface area of the Galaxy
(assumed to be independent of $z$) is $\Sigma$, then $f(z) =
C V(z) / \Sigma = (C \mathcal{N} / \Sigma)
N(z)^{-1}$.  The exponential decay of $N(z)$ must then imply matching
exponential growth in $V(z)$ and in $f(z)$: clouds will be proportionally
larger in regions where the
ambient pressure is lower. This case corresponds to $\alpha = -1$
(i.e., $H_N = H_f$), as claimed for the WIM by BMM06.

We can now generalise the above scenario, and allow $C$ and
$\mathcal{N}$ to both evolve with $z$. If $\alpha = -1$, then
$\mathcal{N}(z) \propto C(z)^{-1}$, i.e., individual clouds at $z=z_1$
might contain more electrons than those at $z=z_2$, but
in this case the total number of clouds at $z_1$ will be smaller than
that at $z_2$ by the same factor.
The total mass of the WIM
remains constant as a function of $z$.

Now consider the case we have derived here, with $\alpha =-0.73$
(and $H_N < H_f$). The fact that $\alpha > -1$ indicates that a cloud
of a given electron content will be larger in a region of low pressure
compared to one under high pressure,
but that the increase in size will be less than that expected from the
ratio of pressures alone.
This could be due to $T_e$ decreasing with $z$, but
this would, if anything, be opposite to the observed trend
\citep{rht99}.  The alternative is that $C(z)\mathcal{N}(z)$ is
decreasing with $z$, i.e., the total particle content
of the ensemble of clouds (and hence the mass of the WIM)
at high $z$ is smaller than 
for those at low $z$.  Quantitatively, we can write $n(z) \equiv f(z)
N(z) = C(z) \mathcal{N}(z) / \Sigma$. Equation~(\ref{eqn_nz})
then implies that $\mathcal{N}(z)  C(z) \propto e^{-z/H_n}$.
We interpret
this as the increasing scarcity of warm gas at higher $z$,
tracing the transition from the WIM into the hot halo.  Electrons
at the surface of WIM clouds are heated to $10^5-10^6$~K via thermal
conduction \citep{cm77,vh07}; they then no longer emit in H$\alpha$,
and have too low a density to be seen in pulsar dispersions (see
\S\ref{sec_hot} below).  Such material is thus no longer observable
through the DMs or EMs considered in \S\ref{sec_analysis}.

The implication of the above discussion is that of the various
parameters 
($H_n$, $H_{n^2}$, $H_N$, $H_f$, $\alpha$, $\beta$) that
we have used to describe the behaviour of the WIM as a
function of $z$, 
the two quantities that best encapsulate the physical conditions
of the gas are $H_N$ and $H_n$.  The former traces the approximate fall in
gas pressure away from mid-plane, as moderated by the gravity of
the disk and by the effects of magnetic and cosmic-ray pressure.
Separately, the latter reveals the extent to which the WIM permeates
into the halo, and so probes the lifetimes of warm clouds embedded
in hot coronal gas. Since these are two separate physical processes,
it is unsurprising that they have distinct scale heights.  The fact
that $H_N \ne H_n$ then demands the existence of a third scale
height, $H_f$, which describes the geometry and size that clouds
adopt as they accommodate external pressure but at the same time
evaporate.

\subsection{Constraints on Hot Coronal Gas}
\label{sec_hot}

At significant distances from the Galactic plane ($z \ga 2000$~pc),
theory predicts that $\ga90\%$ of the ISM by volume should be
occupied by $10^5-10^6$~K coronal gas \citep{fer98b,kbs+99,avi00}.
To interpret our data at these heights above the plane, we first
must determine whether free electrons associated with coronal gas
can contribute significantly to the observed DMs \citep{wmht95,bl07},
or whether we are only probing the WIM.

The standard picture of the Galactic halo is an extended high-$z$
region of very low electron density, $n_{halo} \la
10^{-3}-10^{-4}$~cm$^{-3}$ \citep[e.g.,][]{sws+03,yw05}. We can
test for the presence of such gas in our DM data by attempting a
linear fit of DM$.\sin |b|$ vs. $z$ to the data in Figure~\ref{fig_dmz}
(and relaxing the previous constraint that $|b| \ge 40^\circ$), but
only considering systems for which $z > 5000$~pc, where the
contribution from the WIM should be small.  The best fit slope to
these six measurements is consistent with zero; bootstrapping
provides a $3\sigma$ upper limit on the halo density $n_{halo} <
7.6\times10^{-4}$~cm$^{-3}$. We conclude that any contribution to
pulsar DMs from hot, low density gas in the halo is below our
detection limit.

An independent insight into the relative contributions of the WIM
and corona to pulsar DMs comes from the study of Howk
\etal\ (2003, 2006)\nocite{hss03,hss06}, who compared a variety of
ISM tracers, including pulsar DMs, in the foreground of the globular
cluster NGC~5272 (Messier~3). NGC~5272 contains four radio pulsars,
and corresponds to the blue data point with $z = 10.2$~kpc, DM$.\sin
|b| = 25.9$~pc~cm$^{-3}$ in Figure~\ref{fig_dmz}. \cite{hss06} compute
separate electron column densities along this sight line for the WIM,
for $\sim10^5$~K coronal gas and for $\sim10^6$~K coronal gas. The
total DM inferred from pulsar observations 
is $26.4\pm0.3$~pc~cm$^{-3}$ \citep{hrs+07}, of which
\cite{hss06} estimate that
the WIM contributes $\sim25$~pc~cm$^{-3}$ 
while $10^5$~K gas
contributes $\sim1.5$~pc~cm$^{-3}$.
Any remaining contribution from $10^6$~K gas must be small:
\cite{hss06} adopt a model in which this gas has an average electron
density of $2.4\times10^{-4}$~cm$^{-3}$.  The DM contribution from the
globular cluster itself is expected to be negligible \citep{fkl+01,hss03}.
These results confirm that almost the entire electron column 
out to high $z$ originates in warm gas, at an electron temperature
$\sim10^4$~K. The pulsar DMs in Figure~\ref{fig_dmz} thus predominantly
trace the WIM, and do not constrain the properties of hotter, coronal gas.

\subsection{Warm Ionised Gas in the Galactic Halo}

Given that our data do not appear to be substantially contaminated
by coronal gas, we can now consider what our results imply for the
filling factor and pressure of the WIM at high~$z$.

The physical requirement that $f \le 1$ implies that
Equation~(\ref{eqn_fz}) and the results that follow are not
well-defined for heights $z \ga 2200\pm300$~pc. The value $H_{n^2}
= 400\pm 30$~pc has been derived from H$\alpha$ data in the
range $6^\circ \le |b| \le 35^\circ$ \citep{hrt99}, which at the
distance to the Perseus arm corresponds to a maximum height $z =
1400$~pc.  The EM data used to derive the dependence of $f$ on $z$
thus fall in the regime where $f \le 1$, so that our approach is
internally self-consistent. At the top of this layer, we find $f =
0.30$, $N \approx 0.022$~cm$^{-3}$ and $n \equiv fN \approx 0.0065$~cm$^{-3}$.
This agrees with the findings of BMM06, who argue that $f(z)$ reaches
a maximum value $f \approx 0.3$ at $z \approx 1$~kpc.  However,
this calculation contrasts with previous claims that the WIM is the
dominant phase of the ISM at $z \sim 1$~kpc \cite[e.g.,][]{rey89,rey91b}.
While the WIM probably dominates over neutral gas in these regions,
at this height coronal gas already occupies most of the volume of
the ISM \citep[e.g.,][]{kbs+99}. 

Beyond the top of the WIM layer studied by WHAM in the Perseus arm,
Equation~(\ref{eqn_Nz}) cannot be meaningfully interpreted, since
we no longer have independent constraints on both $n$ and $n^2$.
However, we can broadly characterise the pressure and filling factor
of the WIM at high $z$ by noting that $P \propto N \propto n/f$.
At large $z$, we know that $n$ continues to decay at least as fast
as the exponential in Equation~(\ref{eqn_nz}), because of the high
quality of the fits of this model to the data in Figures~\ref{fig_dmz}
\& \ref{fig_nz} (for $z \ga 5$~kpc, we cannot rule out the possibility
that $n$ begins to fall off more rapidly).  While $f$ appears to
increase with distance from the mid-plane at low values of $z$,
this function must turn over at $z \ga 2000$~pc to accommodate
coronal gas \citep[e.g.,][]{kbs+99,avi00}.  If $n$ and $f$ are both
decreasing functions at high $z$, then $N \equiv n/f$ at these
heights must become increasingly larger than predicted by
Equation~(\ref{eqn_Nz}).  This then implies that any warm ionised
gas at high $z$ is over-pressured relative to the exponential decay
in pressure seen within 2~kpc of the mid-plane. This high pressure
is consistent with the expectation that any small warm clouds at
these heights should be enveloped in a hot halo \citep[e.g.,][]{fws+05}.

At $z\approx5-10$~kpc, the ambient pressure in the Galactic halo
is $P/k \approx 1000-2000$~K~cm$^{-3}$ \citep{ss94b,smw02,ezb+08}.
If we assume $T_e \approx 10^4$~K for the WIM, and adopt
Equation~(\ref{eqn_nz}) as an upper limit on $n$ at these values
of $z$, we infer $f < 0.02$ for any warm ionised gas at these heights
above the plane.  These small cloudlets may correspond to the
``drizzle'' of ionised gas accreting into the halo proposed by
\cite{bsam07}, or to the extraplanar droplets and clumps of H$\alpha$
seen in observations of some edge-on spiral galaxies \citep{fwg96,cbvf01}.

\bigskip
\bigskip

\section{Conclusion}

We have used the dispersion measures of pulsars at known, reliable, distances
and the emission measures traced by diffuse H$\alpha$ emission
to re-evaluate the structure and distribution of the thick disk
of warm ionised gas in the Milky Way.
Our main conclusions are as follows:
\begin{enumerate}

\item The DMs of high-latitude pulsars are well fit by a plane-parallel
distribution of free electrons, in which the volume-averaged electron
density, $n$, decays exponentially with height, $z$, above the
Galactic plane.

\item Within a few kpc of the Sun, the exponential scale height for
$n$ is $H_n\approx1800$~pc, approximately double the value
determined in most previous studies.  Estimates of distances to high-latitude
pulsars using the ``NE2001'' electron model of Cordes \& Lazio
(2002, 2003\nocite{cl02,cl03}) are thus likely to be substantial
underestimates.

\item Within a few kpc of the Sun, the mid-plane volume-averaged
electron density of the diffuse WIM is $n_0 = 0.014\pm0.001$~cm$^{-3}$,
and the corresponding total vertical column density of the WIM is
$\approx26$~pc~cm$^{-3}$.  Assuming a total vertical emission measure
EM$_0 = 1.9\pm0.3$~pc~cm$^{-3}$, clouds in the thick-disk
component of the WIM have a typical
internal electron density of $N_0 = 0.34\pm0.06$~cm$^{-3}$ at mid-plane, with a volume
filling factor $f_0 = 0.04\pm0.01$.

\item Comparison of DMs and EMs shows that the internal electron
density, $N$, of individual ionised clouds within the thick-disk
WIM decays rapidly with $z$, with an exponential scale height $H_N
\approx 500$~pc.  The differing behaviour for $H_n$ and $H_N$
requires that the filling factor of the WIM grows exponentially
with $z$ over the range $0 \le z \la 1.4$~kpc, with a scale height
$H_f \approx 700$~pc.

\item The filling factor of the thick-disk WIM probably reaches a
maximum of $f\approx 0.3$ at $z \approx 1.5$~kpc. At larger distances
from the Galactic plane, the ISM becomes increasingly dominated by
hotter, coronal, gas in the halo.  However, hot low-density halo
gas makes only a negligible contribution to observed pulsar DMs.

\end{enumerate}

A variety of future observations will allow us to test the above
conclusions, and to better understand the structure of the diffuse WIM.
Most immediately, from late 2008 WHAM will be relocated to the
southern hemisphere, allowing studies similar to those 
carried out on the Perseus spiral arm to now be repeated
for other regions.

Further improvements will come from an increase in the number of
high-latitude radio pulsars with direct distance estimates.  For example,
only $\approx25\%$ of the Galactic globular clusters at $|b| \ge 40^\circ$
contain known radio pulsars.  While some clusters at these latitudes have
been searched for pulsations without success \citep{and93,bl96,hrs+07},
at least some of these targets probably contain as yet undetected pulsars
\citep[e.g., NGC~5634;][]{shq02}, and should be searched more deeply.
Furthermore, there are several high-latitude pulsars with 1.4~GHz radio
fluxes at the level of a few mJy, which is sufficiently bright to derive
interferometric parallaxes through future observations.

Most surveys for radio pulsars
have concentrated on the Galactic plane, and those surveys that have
covered high-latitude regions have been of comparatively modest
sensitivity \citep[e.g.,][]{mld+96}.  Forthcoming all-sky pulsar
surveys with more sensitive instrumentation will greatly expand the
sample of high-$|b|$ pulsars, whose DMs can then be used to better
map out the structure of the WIM. 

Finally, we note that the new model we have presented for the WIM
can also be used to probe Galactic magnetic fields at high $z$.  In
a subsequent paper, we will combine the density distribution derived
here with Faraday rotation measures of a large number of polarised
extragalactic sources at high $|b|$, with which we aim to derive
the strength and geometry of magnetic fields in the Galactic halo.

\section*{Acknowledgments} 

We thank Bob Benjamin, Joss Bland-Hawthorn, Jim Cordes, Ralf-J\"urgen
Dettmar, Craig Heinke, Jeff Hester, Adam Leroy, Jay Lockman, Naomi
McClure-Griffiths, Dick Manchester and Stephen Ng for useful
discussions.  We also thank the anonymous referee for helpful
comments that led to the addition of \S\ref{sec_sep} to the paper.
B.M.G.\ acknowledges the support of a Federation Fellowship from
the Australian Research Council through grant FF0561298.  G.J.M.\
acknowledges the support of the National Science Foundation through
grant AST-0607512.

\bibliographystyle{apj}
\bibliography{journals,modrefs,psrrefs,crossrefs}

\clearpage

\begin{table}
\caption{Estimates of the scale height and mid-plane value
of the volume-averaged
electron density in the thick-disk component of the WIM. The entries have
been sorted by increasing scale height.}
\label{tab_n0}
\medskip
\begin{tabular}{ccll} \hline
$H_n$ (pc) & $n_0$ (cm$^{-3}$) & Reference & Comment\\ \hline
$430-670$ & $0.028-0.035$ & \cite{rd75} & \\
$670^{+170}_{-140}$ & $0.033\pm0.002$ &  \cite{nct92b} & \\
$83$0 & $0.025\pm0.005$ & \cite{pw02} \\
$880\pm60$ & $0.019$ & \cite{tc93} & Thick-disk component of two-component model \\
$890^{+250}_{-180}$ & $0.036$ &  \cite{sed90} & \\
$900$ & $0.025$ & \cite{rey91b} & Thick-disk component of two-component model \\
$930\pm130$ & $0.023\pm0.004$ & \cite{bm08} & \\
$950$ & $0.035$ & \cite{cl02} & Thick-disk component of two-component model \\
$>1000$ & $0.025$ & \cite{lmt85} & Thick-disk component of two-component model \\
$>1000$ & $0.021\pm0.001$ &  BMM06 & \\ 
$1070-1100$ & $0.018-0.020$ & \cite{gbc01} & Thick-disk component of two-component model \\
$1830^{+120}_{-250}$ & $0.014\pm0.001$ & This paper & Only uses data at $|b| \ge 40^\circ$ \\ \hline
\end{tabular}
\end{table}

\begin{table}
\caption{Estimates of the mid-plane volume filling factor
of the WIM (sorted by decreasing $f_0$).}
\label{tab_f0}
\medskip
\begin{tabular}{cl} \hline
$f_0$ & Reference \\ \hline
$0.2$ & \cite{cwf+91} \\
$0.11$  & \cite{pw02} \\
$0.1-0.5$ & \cite{pyn93} \\
$0.1-0.2$ & \cite{rey91,rey97} \\
$0.1$ & \cite{kh87,kh88b} \\
$0.083$ & \cite{cox05} \\ 
$0.05-0.14$ & \cite{fer01} \\
$0.08\pm0.02$ & \cite{bm08} \\
$0.05\pm0.01$ & BMM06 \\
$0.04\pm0.01$ & This paper \\ \hline
\end{tabular}
\end{table}

\begin{table}
\caption{Estimates of the mid-plane mean square electron density 
of the WIM (sorted by decreasing $(n^2)_0$).}
\label{tab_n20}
\medskip
\begin{tabular}{cl} \hline
$(n^2)_0$ (cm$^{-6}$) & Reference \\ \hline
$0.009$ & \cite{kh87} \\
$0.008\pm0.001$ & BMM06\nocite{bmm06} \\
$0.007$ & \cite{rey90c} \\
$0.005-0.012$ & \cite{fer01} \\
$0.005\pm0.001$ & \cite{bm08} \\
$0.005\pm0.001$ & This paper \\ \hline
\end{tabular}
\end{table}

\end{document}